\documentclass[aps,prl,showpacs,twocolumn,superscriptaddress]{revtex4}
\usepackage{amsmath}
\usepackage{graphicx,color,epsfig}

\begin{document}

\title{Generalized synchronization of chaos in autonomous systems}
\author{O. Alvarez-Llamoza}
 \affiliation{Departamento de F\'{\i}sica, FACYT, Universidad de
        Carabobo, Valencia, Venezuela}
\affiliation{Centro de F\'isica Fundamental, Universidad de Los Andes, M\'erida, M\'erida 5251, Venezuela}
\author{M. G. Cosenza}
\affiliation{Centro de F\'isica Fundamental, Universidad de Los Andes, M\'erida, M\'erida 5251, Venezuela}
\date{\today}

\begin{abstract} 
We extend the concept of generalized synchronization of chaos, a phenomenon that occurs in driven dynamical systems, to the context of autonomous spatiotemporal systems. It means a situation where the 
chaotic state variables in an autonomous system can be synchronized to each other 
but not to a coupling function defined from them. 
The form of the coupling function is not crucial; it may not depend on all the 
state variables 
nor it needs to be active for all times for achieving generalized synchronization.
The procedure is based on the analogy between a response map subject to an external drive acting with a probability $p$ and an autonomous system of coupled maps where a global interaction between the maps takes place with this same probability. 
It is shown that, under some circumstances, the conditions for stability of
generalized synchronized states are equivalent in both types of systems. 
Our results reveal the existence of similar minimal conditions for the emergence of generalized synchronization of chaos in driven and in autonomous spatiotemporal systems. 
\end{abstract}

\maketitle
Generalized synchronization of chaos is a common phenomenon occurring in 
unidirectionally coupled systems, where a distinction can be made between 
a drive or forcing subsystem and another driven or response subsystem \cite{Pikovsky}. It arises when a functional relation different from the identity is established between the drive and the response subsystems \cite{Rulkov1}. This phenomenon has been the subject of many theoretical and numerical studies \cite{Rulkov2,Kapitaniak,Hunt,Parlitz99,Roy} and has been observed experimentally \cite{Peterman,Gauthier,Rulkov3,Tang}.
On the other hand, there has been recent interest in the investigation of chaotic synchronization and other collective behaviors emerging in networks of mutually interacting dynamical elements where no external influences are present \cite{Newman,Manrubia}. 
In particular, the phenomenon of complete synchronization, where all the state variables converge to a single trajectory in phase space, has been widely studied in these autonomous dynamical systems. 
In this paper, we present a procedure that allows to extent the concept
of generalized synchronization of chaos found in driven systems to the context of autonomous systems.
We mean that, under some circumstances,
the chaotic state variables in an autonomous dynamical system can be synchronized to each other
but not to a coupling function containing partial information from those variables. This phenomenon can be seen as a form of collective behavior arising in some specifically designed dynamical networks.
Our procedure is based on the reported analogy between a single driven map and a system of globally coupled maps \cite{Us1}. This analogy provided an explanation of dynamical clustering \cite{Us2,Manrubia} and of stability of steady states in 
systems with delayed interactions \cite{Marti}. Here we search for minimal conditions for the occurrence of generalized synchronization of chaos in both, driven and autonomous systems, by using models of coupled maps. 

We consider a map driven with a probability $p$,
\begin{equation}
\begin{array}{l}
x_{t+1} = \begin{cases}
w(x_t,y_t), & \text{with probability $p$}\\
f(x_t),  & \text{with probability $(1-p)$},
          \end{cases} \\
y_{t+1} =  g(y_t),
\end{array}
\label{2dmap}
\end{equation}
where $f(x_t)$ and $g(y_t)$ describe the dynamics of the driven and the drive variables, respectively; and
the coupling relation between them is chosen 
to be  
\begin{equation}
  w(x_t,y_t)=(1-\epsilon)f(x_t)+ \epsilon g(y_t) \, ,
\end{equation}
where $\epsilon$ is the coupling strength.
For the driven chaotic dynamics we shall choose in most examples $f(x_t)=b+ \ln | x_t |$,
where $b$ is a real parameter and $x_t \in (-\infty, \infty)$. This logarithmic
map exhibits robust chaos, with no periodic windows and no separated chaotic
bands, on the interval $b \in [-1,1]$ \cite{kawabe}.

The linear stability condition for generalized synchronization is determined by the Lyapunov exponents of the
two-dimensional system Eq.~(\ref{2dmap}). These are defined as $\Lambda_x = \lim_{T\rightarrow \infty} \ln L_x$ and $\Lambda_y = \lim_{T\rightarrow \infty} \ln L_y$, where $L_x$ and $L_y$ are the magnitude of the eigenvalues of $[\prod^{T-1}_{t=0} \mathbf{J}(x_t,y_t)] ^{1/T}$, and $\mathbf{J}(x_t,y_t)$ is the Jacobian matrix for the system  Eq.(\ref{2dmap}), calculated along an orbit. A given orbit $\{x_t,y_t\}$ from $t=0$ to $t=T-1$ can be separated in two subsets,
according to the source of the $x_t$ variable, either coupled or uncoupled, that we respectively denote as 
$A=\{ \{x_t,y_t\}:  x_t=w(x_{t-1},y_{t-1}) \}$ possessing $pT$ elements, and $B=\{ \{x_t,y_t\}: x_t=f(x_{t-1}) \}$ having $(1-p)T$ elements. We get
\begin{small}
\begin{equation}
\left( \prod^{T-1}_{t=0} \mathbf{J}\right)^{1/T} = \left( \begin{array}{cc}
\displaystyle{ \prod_{t: \, x_t \in A} w_x \prod_{t: \, x_t \in B} f'(x_t)} & K\\
0 & \displaystyle{\prod^{T-1}_{t=0} g'(y_t)}
\end{array} \right)^{1/T},
\label{jacobians2}
\end{equation}
\end{small}where $w_x=\frac{\partial w}{\partial x}=(1-\epsilon)f'(x)$, and $K$ is a polynomial whose terms contain products of $w_x$, $\epsilon$, and $g'(y_t)$ to be evaluated along time.
Then $L_x=[ \prod_{x_t \in A} w_x \prod_{x_t \in B} f'(x_t)]^{1/T}$
and $L_y=[ \prod^{T-1}_{t=0} g'(y_t) ]^{1/T}$. Thus we get
\begin{small}
\begin{equation}
\Lambda_x= p \ln|1-\epsilon| + \lim_{T \rightarrow \infty} \dfrac{1}{T}  \left[ \ln \prod_{x_t \in A} \vert f'(x_t)\vert + \ln  \prod_{x_t \in B} \vert f'(x_t)\vert \right]
\end{equation}
\begin{equation}
 \Lambda_y =\lim_{T \rightarrow \infty} \dfrac{1}{T} \sum^{T-1}_{t=0} \ln | g'(y_t) | = \lambda_g \, ,
\end{equation}
\end{small}where $\lambda_g$ is the Lyapunov exponent of the map $g(y_t)$. Generalized synchronization  occurs if the Lyapunov exponent corresponding to the driven map is negative \cite{Rulkov1}; i.e., $\Lambda_x<0$. 

\begin{figure}[h]
\centerline{
\includegraphics[scale=0.8]{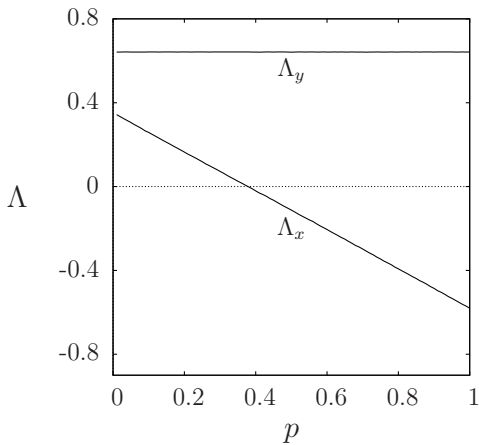}}
\caption{Lyapunov exponents $\Lambda_y$ and $\Lambda_x$ for the driven system Eq.(\ref{2dmap}) as a function of $p$, for  $g(y_t)=0.5 + \ln | y_t |$ and $f(x_t)=-0.7+ \ln |x_t|$, and $\epsilon=0.7$.}
\label{fig0}
\end{figure}

Figure~\ref{fig0} shows the Lyapunov exponents of the driven system Eq.~(\ref{2dmap}) with $g \neq f$, when the probability $p$ is varied. The exponent $\Lambda_y$ is constant and positive for the chosen chaotic drive $g$. On the other hand, there is a definite value of the $p$ at which the exponent $\Lambda_x$ changes its sign, from positive to negative, signaling the onset of generalized synchronization and the appearance of a contracting direction in the dynamics of the two-dimensional map Eq.~(\ref{2dmap}). 

\begin{figure}[t]
\centerline{
\includegraphics[scale=0.55]{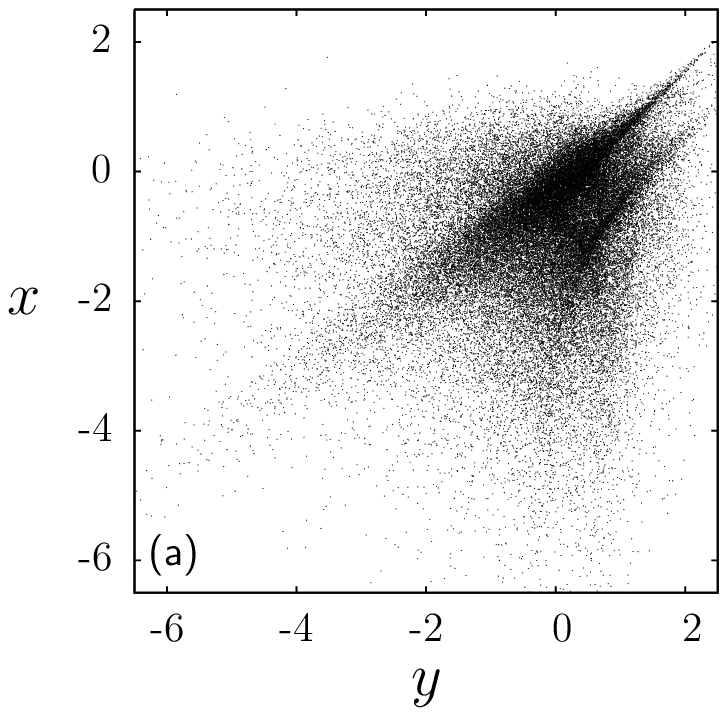}
\hspace{2mm}
\includegraphics[scale=0.55]{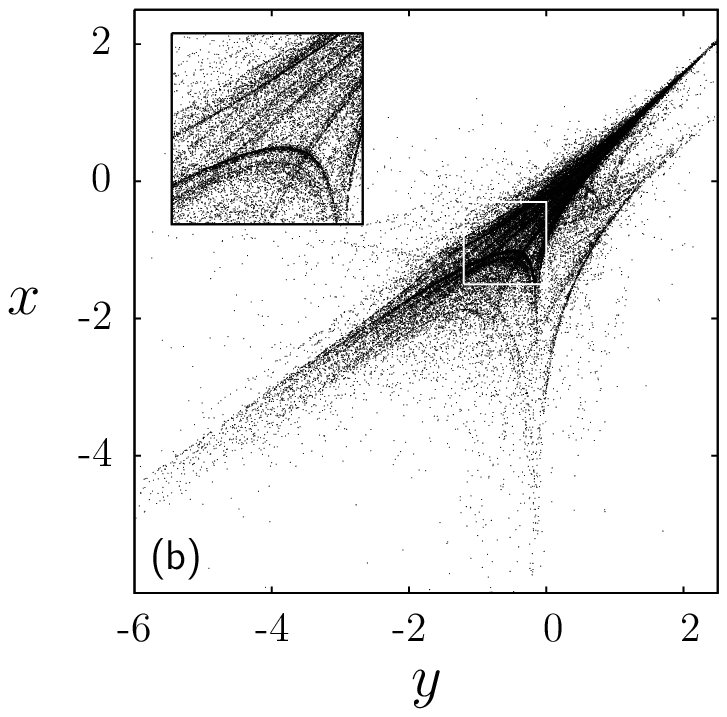}}
\caption{Orbits of the driven map Eq.(\ref{2dmap}) for  $g(y_t)=0.5 + \ln | y_t |$ and $f(x_t)=-0.7+ \ln |x_t|$, $\epsilon=0.7$. 
(a) $p=0.3$ (unsynchronized). (b) $p=0.9$ (generalized synchronization); the insert is a magnification of the marked square.}
\label{fig1}
\end{figure}

Figure~\ref{fig1} shows the orbits of the driven system Eq.(\ref{2dmap}) 
for different values of the probability $p$. Generalized synchronization in this two-dimensional map system is manifested by the appearance of a strange attractor for values of $p$ above some threshold value.

\begin{figure}[h]
\centerline{
\includegraphics[scale=0.8]{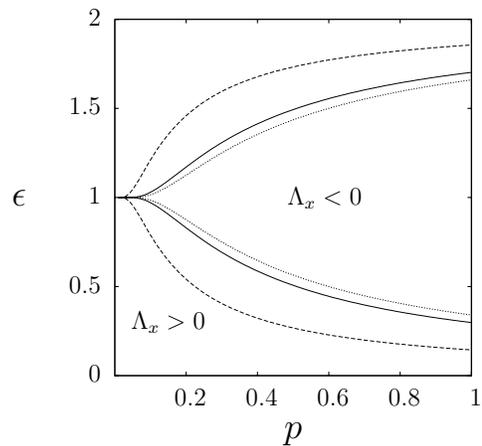}}
\caption{Boundaries $\Lambda_x = 0$ (Eq.~(\ref{lyapexp})) on the plane $(p,\epsilon)$ for complete synchronization in
the driven map Eq.~(\ref{2dmap}) with $f(x_t)=-0.7+ \ln |x_t|$. Dashed line: $g(y_t)=\{ \overline{x}_1=-0.855762\}$; solid line: $g(y_t)=-0.7+\ln|y_t|$ (this boundary also corresponds to complete synchronization in the autonomous system Eqs.~(\ref{auto_syst}) and  (\ref{partial})); dotted line: $g(y_t)=\{ \overline{x}_1=0.18049, \overline{x}_2=-2.41208 \}$.}
\label{fig2}
\end{figure}

When $g=f$, the condition $\Lambda_x<0$ implies
complete synchronization, where $x_t=y_t$. In this case we get 
\begin{equation}
\Lambda_x= p \ln |1-\epsilon| + \lambda_f \, ,
\label{lyapexp}
\end{equation}
where $\lambda_f$ is the Lyapunov exponent of the map $f$.
Figure~\ref{fig2} shows the stability boundaries, given by $\Lambda_x=0$, for the completely synchronized 
states of the system  Eq.~(\ref{2dmap}) on the space of parameters $(p,\epsilon)$
for different orbits of a drive $g(y_t)$.
When $g=f$ complete synchronization in a unstable periodic-$m$ orbit of the
map $f$, defined by $f^{(m)}(\overline{x}_n)=\overline{x}_n$ and satisfying $e^{\lambda_f}=\prod^m_{n=1}
|f'(\overline{x}_n)|>1$, where
$\{\overline{x}_1,\overline{x}_2,\ldots,\overline{x}_m\}$ are the set of
consecutive points on this orbit, can also be achieved in the system Eq.~(\ref{2dmap}), as shown in Fig.~\ref{fig2}. 

\begin{figure}[t]
\centerline{
\includegraphics[scale=0.8]{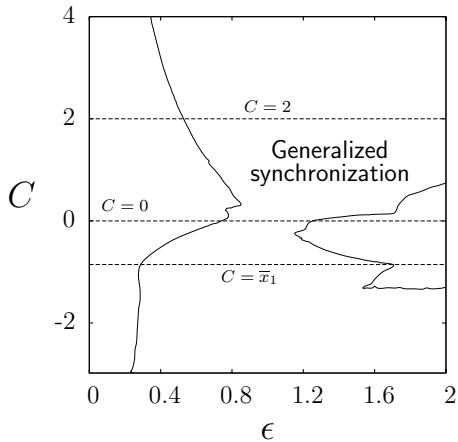}}
\caption{Region for generalized synchronized states $x_t \neq C$ (satisfying $\Lambda_x<0$) for the system Eq.~(\ref{2dmap})
subject to a 
constant drive $g(y_t)=C$, on the plane $(C,\epsilon)$. Horizontal lines indicate the values $C=2$, $C=0$, and $C=\{\overline{x}_1=-0.855762\}$. 
The synchronization region for $C\geq 0$ also comprises the generalized synchronized states ($\langle \sigma \rangle < 10^{-7}$) of the autonomous system Eq.~(\ref{auto_syst}) with $N=10^4$, having a coupling function given by Eq.~(\ref{cte}). In both cases,  $f(x)=-0.7+ \ln |x|$ and $p=0.5$.}
\label{fig3}
\end{figure}

On the other hand, if $g\neq f$, the condition $\Lambda_x<0$  corresponds to generalized synchronization, characterized by $x_t \neq y_t$. Consider, for example, 
the system Eq.~(\ref{2dmap}) subject to an intermittently applied, constant drive $g(y)=C$, which reduces to a one-dimensional map with a Lyapunov exponent  $\Lambda_x$ depending on the parameters $C$, $\epsilon$, and $p$. The region where generalized synchronization arises on the plane $(C,\epsilon)$ for a fixed value of $p$ is shown in Fig.~\ref{fig3}.

The dynamics of the single driven map  Eq.~(\ref{2dmap}) can be compared
with the dynamics of an autonomous system of maps that share a global interaction with a probability $p$, $\forall i$,
\begin{equation}
x^i_{t+1} = 
\begin{cases}
(1-\epsilon)f(x^i_t)+ \epsilon H_t( x^j_t: j \in Q_t), ~ \text{with prob. $p$},\\
f(x_t^i), \hspace{3.2cm} \text{with prob. $(1-p)$,} &
\end{cases}
\label{auto_syst}
\end{equation}
where $x^i_t$ ($i=1,2,\ldots,N$) gives the state of the $i$th
map at discrete time $t$, $\epsilon$ is the strength of the coupling to the global interaction function $H$; and $Q_t$ is a subset having $q \leq N$ elements of the system that may be chosen at random at each time $t$. 
Each map receives the same information from the coupling function $H$ at any $t$ with probability $p$.
When this autonomous system gets synchronized at some values of parameters, we have $x_t^i=x_t$. Thus for this same set of parameters, the single driven map subject to a forcing that satisfies $g(y_t)=H(x^j_t=x_t: j \in Q_t)$ for long times should exhibit a synchronized state similar to that of the associated autonomous system Eq.~(\ref{auto_syst}).
Thus, besides complete synchronization where $x_t^i=x_t=H$, 
other synchronized states, characterized by $x_t^i=x_t \neq H$ and which we call generalized synchronization, should also occur in the autonomous system Eq.~(\ref{auto_syst}) for appropriate values of parameters. 

A synchronized state  can be characterized by the
asymptotic time-average $\langle \sigma \rangle$ (after discarding a number of transients) of the instantaneous standard deviations
$\sigma_t$ of the distribution of map variables $x^i_t$, defined as
\begin{equation}
\sigma_t=\left[ \frac{1}{N} \sum_{i=1}^N \left( x^i_t - \langle x_t \rangle \right)^2 \right]^{1/2},
\end{equation}
where $\langle x_t \rangle$ is the instantaneous mean of the values $x^i_t$, $\forall i$. Stable synchronization corresponds to $\langle \sigma \rangle=0$. Here we use the numerical criterion $\langle \sigma \rangle < 10^{-7}$.

As an example of complete synchronization, consider a partial mean field coupling function
\begin{equation}
\label{partial}
 H=\frac{1}{q} \sum^q_{j=1} f(x^j_t).
\end{equation}

In this case the autonomous system Eq.~(\ref{auto_syst}) can be expressed in vector form as
\begin{equation}
 \mathbf{x}_{t+1} = 
   \begin{cases}
  \left( (1-\epsilon) \mathbf{I} + \dfrac{\epsilon}{q}\mathbf{G}_t \right)  \mathbf{f}(\mathbf{x}_t),
        \quad \text{with probability $p$},\\
   \mathbf{I} ~ \mathbf{f}(\mathbf{x}_t), \hspace{2.1cm} \text{with probability $(1-p)$},
   \end{cases}
\label{aut_syst_vect}
\end{equation}
where the $N$-dimensional vectors $\mathbf{x}_t$ and $\mathbf{f}(\mathbf{x}_t)$
have components $[\mathbf{x}_t]_i=x_t^i$ and
$[\mathbf{f}(\mathbf{x}_t)]_i=f(x_t^i)$, respectively, 
$\mathbf{I}$ is the $N \times N$ identity matrix, and $\mathbf{G}_t$ is an $N\times N$ matrix 
that at each time $t$ possesses $q$ randomly chosen columns that have all their components equal to $1$ while the remaining $N-q$ columns have all their components equal to $0$.
The case $q=N$ and $p=1$ corresponds to the usual mean field global coupling \cite{Kaneko2}. The linear stability analysis \cite{Waller} of the complete synchronized state $f(x_t^i)=f(x_t)=H$ yields
\begin{equation}
\label{condition}
 \left| \left[ (1-\epsilon) + \frac{\epsilon}{q}\alpha_k \right]^p  e^{\lambda_f} \right| < 1 \, ,
\end{equation}
where $\alpha_k=\delta_{0k}q \,$ $(k=0,1\ldots,N-1)$ are the set of eigenvalues of the matrix $\mathbf{G}_t$ for any $t$, with the zero eigenvalue having $(N-1)$-fold degeneracy. The eigenvector corresponding to $k=0$ is homogeneous.
Thus only perturbations of $\mathbf{x}_t$ along the other 
eigenvectors may destroy the coherence.
Thus, condition Eq.~(\ref{condition}) with $k\neq 0$
becomes
\begin{equation}
p \ln |1-\epsilon |+ \lambda_f < 0,
\end{equation}
which is the same condition for stability of complete synchronized states in the driven map, Eq.~(\ref{lyapexp}), when $g=f$. Thus the boundary that separates the region where complete synchronization occurs on the space of parameters $(p,\epsilon)$ for the autonomous system Eq.~(\ref{auto_syst}), with $H$ given by Eq.~(\ref{partial}) for any value of $q$, coincides with the stability boundary $\Lambda_x=0$ in Fig.~\ref{fig2} for the driven system with $g=f$. However, in contrast with the driven case, the unstable periodic orbits of the local map $f$ cannot be synchronized in the autonomous system because they correspond to unstable synchronized states in this system. 

For other functional forms of the coupling function $H$ it is possible to find generalized synchronized states in the autonomous system Eq.~(\ref{auto_syst}).
For example, consider the coupling function
\begin{equation}
\label{cte}
 H= \frac{\sum^q_{j=1} f(x^j_t)}{\sum^r_{j=1} f(x^j_t)} \, ,
\end{equation}
where $q<N$, $r<N$, and define the instantaneous mean field of the system as
\begin{equation}
\label{mean}
 S_t(N)=\frac{1}{N} \sum^N_{j=1} f(x^j_t) \, .
\end{equation}
\begin{figure}[t]
\centerline{
\includegraphics[scale=0.8]{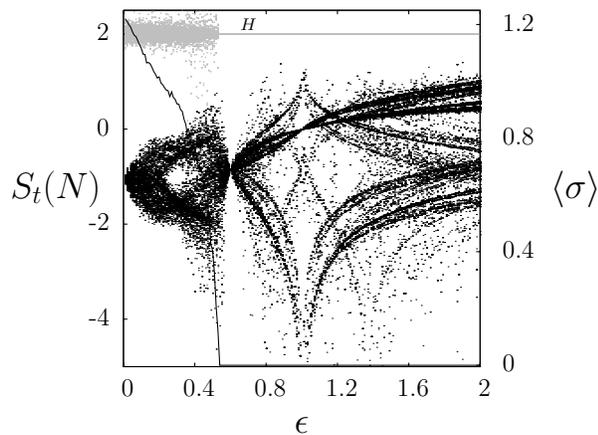}} 
\caption{Left vertical axis: Bifurcation diagrams of $S_t(N)$ (black dots) and $H$ (grey dots)  
vs. $\epsilon$ for the autonomous system given by Eqs.~(\ref{auto_syst}) and~(\ref{cte}) with $f(x_t^i)=-0.7+ \ln |x_t^i|$. Right vertical axis: $\langle \sigma \rangle$ vs. $\epsilon$ (continuous line). Fixed parameters: $p=0.5$, $N=10^4$.}
\label{fig6}
\end{figure}
Figure~(\ref{fig6}) shows the bifurcation diagram of $S_t(N)$
as well as $\langle \sigma \rangle$ as functions of the coupling parameter $\epsilon$ for the autonomous system with the coupling function $H$ given by Eq~(\ref{cte}).
The mean field $S_t(N)$ is chaotic for all values of $\epsilon$. 
For $\epsilon \geq 0.54$, we obtain $\langle \sigma \rangle =0$, indicating that the system is synchronized in a chaotic state
for that range of the coupling parameter. Figure~(\ref{fig6}) also shows the bifurcation diagram of the function  
$H$ with values of $q$ and $r$ chosen such that $q/r=2$. Note that $H=2$ on the range where synchronization occurs, otherwise it is chaotic. This happens independently of the specific values of $q$ and $r$, as long as $q/r=2$. Thus, we have a situation where dynamical elements in an autonomous system converge to a chaotic synchronized state $x_t^i=x_t$, while a  coupling function that contains partial information about the system reaches a value different from that state, i.e. $H \neq x_t$. 
This is the analogous to the phenomenon of generalized synchronization of chaos observed in a driven 
system Eq.~(\ref{2dmap}) with constant drive $g=C=2$, as illustrated in Fig.~\ref{fig3}. 
Furthermore, for all given values of $q$ and $r$ the autonomous system with $H$ described by Eq~(\ref{cte}) exhibits generalized synchronization that yields $H=q/r$ on the same region of the plane $(C,\epsilon)$ in Fig.~(\ref{fig3}) where generalized synchronization is observed for the map driven with constant $g=C=q/r>0$. 

\begin{figure}[h]
\centerline{
\includegraphics[scale=0.8]{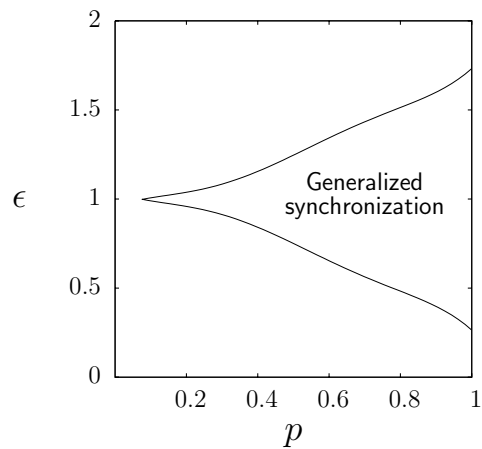}}
\caption{Region of generalized synchronization on the plane $(p,\epsilon)$ for: 
(i) the single driven map Eq.~(\ref{2dmap}) with $g=0$ ($\Lambda_x < 0$); and (ii) the autonomous system Eq.~(\ref{auto_syst}) with $N=10^4$ and $H$ given by Eq.~(\ref{disper}) for any value of $q$ ($\langle \sigma \rangle< 10^{-7}$). In both cases, $b=-0.7$.}
\label{fig7}
\end{figure}

The equivalence between an external drive and a global coupling function can be used to predict the emergence of generalized synchronization in either system from the occurrence of this phenomenon in the other. As an illustration, consider 
a parametrically, intermittently driven map Eq.~(\ref{2dmap}) with $g=0$.  Figure~\ref{fig7} shows the region where 
$\Lambda_x<0$ 
on the plane $(p,\epsilon)$ for this system.
The analogy external drive-global coupling suggests that  
an autonomous spatiotemporal system Eq.~(\ref{auto_syst}) having a coupling function that reached a value $H=0$ at synchronization, should possess the same region of stability for this state on the plane $(p,\epsilon)$. 
Consider, for example, the coupling function
\begin{equation}
H=\left[ \frac{1}{q} \sum_{j \in Q_t}^q \left(f(x^j_t) - S_t(q) \right)^2 \right]^{1/2} \, ,
\label{disper}
\end{equation}
where $S_t(q)$ is the partial mean field of $q$ maps randomly chosen at each time $t$. 
This autonomous system 
reaches generalized synchronization, i.e., a chaotic synchronized state with $x_t^i\neq H=0$ for any value of $q$, on the same region of the plane $(p,\epsilon)$ as in the driven map Eq.~(\ref{2dmap}) shown in Fig.~\ref{fig7}. 

Note that $H$ does not need to be a function of all the maps nor it must be active for all times;
what matters for synchronization is that all elements in the autonomous system share the same minimal information at any time. 
Thus, the nature of the common input (external or endogenous) being received by the local units is irrelevant for synchronization in either driven or autonomous systems. In both scenarios, each unit evolves as a driven map at the local level. 

\begin{figure}[h]
\centerline{
\includegraphics[scale=0.8]{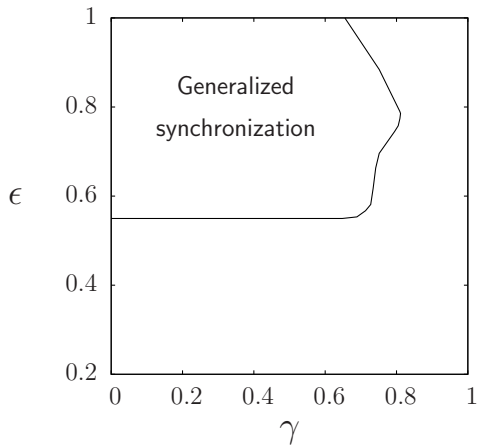}}
\caption{Region for generalized synchronization on the plane $(\epsilon,\gamma)$ for both,
the autonomous network Eq.~(\ref{auto_net}) with $H$ given by Eq.~(\ref{cte}) with $q=200,r=500$; and the  driven network Eq.~(\ref{driven_net}) with constant $g=0.4$. For both systems, $f(x)=4x(1-x)$, $N=10^3$, $p=0.5$.}
\label{fn}
\end{figure}

Generalized synchronization may also arise in an autonomous network of coupled oscillators that, in addition, share a global coupling function $H$ with probability $p$. As an illustration, consider a one-dimensional lattice,
\begin{equation}
x^i_{t+1} = \left\lbrace 
\begin{array}{l}
(1-\epsilon-\gamma)f(x^i_t)+ \frac{\gamma}{2}\left[ f(x_t^{i+1})+f(x_t^{i-1}) \right] + \\ 
   \qquad \qquad \epsilon H( x^j_t:j \in Q_t), \qquad \text{with prob. $p$} ;\\
(1-\gamma)f(x^i_t)+ \frac{\gamma}{2}\left[ f(x_t^{i+1})+f(x_t^{i-1}) \right], \\ 
  \qquad \qquad \text{with probability $(1-p)$}; 
\end{array}
\right. 
\label{auto_net}
\end{equation}
where $\gamma$ is the local coupling parameter. The analogy 
can be established with a lattice of similar coupled maps subject to a uniform drive $g(y_t)$ with probability $p$,
\begin{equation}
x^i_{t+1} = \left\lbrace 
\begin{array}{l}
(1-\epsilon-\gamma)f(x^i_t)+ \frac{\gamma}{2}\left[ f(x_t^{i+1})+f(x_t^{i-1}) \right] + \\
  \qquad \qquad \epsilon g(y_t), \qquad \text{with probability $p$};\\
(1-\gamma)f(x^i_t)+ \frac{\gamma}{2}\left[ f(x_t^{i+1})+f(x_t^{i-1}) \right],\\ 
\qquad \qquad \text{with probability $(1-p)$}. 
\end{array}
\right. 
\label{driven_net}
\end{equation}
Periodic boundary conditions are assumed for both systems. Figure \ref{fn} shows, for a given example, that the region of generalized synchronization is the same for both systems on the plane $(\epsilon,\gamma)$, when $g(y_t)=H$.

In summary, based on the analogy between a single driven map and a globally coupled system of maps, 
we have extended the concept of generalized synchronization of chaos to the context of autonomous dynamical systems.
It means that there can exist a coupling function $H$ containing {\it some} information about the elements in the autonomous system that reaches a state different from the state of those elements when they are chaotically synchronized.
The functional form of $H$ is not crucial for achieving generalized synchronization; what matters is the sharing of the same information about the system by its elements. By comparing a single map driven with a probability $p$ and an autonomous system of maps sharing a global coupling with this same probability, we have shown that the minimal conditions for stability of generalized synchronized states are equivalent in both types of systems.  
The analogy external drive-global coupling allows to design a spatiotemporal autonomous system exhibiting generalized synchronization of chaos from the knowledge of the occurrence of this phenomenon in an associated single driven map. Extensions of this work include the possibility of designing global coupling functions in autonomous systems to achieve specific behaviors or patterns as self-organizing phenomena.

This work was supported by grant C-1579-08-05-B from Consejo de Desarrollo Cient\'{\i}fico, Human\'{\i}stico y Tecnol\'ogico, Universidad de Los Andes, Venezuela.

\end{document}